\documentclass[10pt,reqno]{amsart}

\usepackage{latexsym}
\usepackage{amssymb}
\usepackage{amsmath}
\usepackage{amsthm}
\usepackage{amsfonts}
\usepackage{amssymb}

\numberwithin{equation}{section}

\newcommand{\cDD}{{\mathcal D}}
\newcommand{\cHH}{{\mathcal H}}

\newcommand{\de}{\partial}

\newcommand{\nat}{{\Delta_{\theta}}}
\newcommand{\nar}{{\Delta_{r}}}
\newcommand{\snat}{\sqrt{\Delta_{\theta}}}
\newcommand{\snar}{\sqrt{\Delta_{r}}}
\newcommand{\diag}{{\rm diag}}
\newcommand{\rc}{r_c}
\newcommand{\rp}{r_+}
\newcommand{\rmen}{r_{-}}
\newcommand{\RR}{{\mathbb R}}

\newcommand{\eqn}{\begin{eqnarray}}
\newcommand{\feqn}{\end{eqnarray}}
\newtheorem{theorem}{Theorem}
\newtheorem{lemma}{Lemma}

\begin{document}

\title[Absence of  Time-periodic Solutions for The Dirac Equation in KN-dS Black Hole Background]
{Absence of Normalizable Time-periodic Solutions for The Dirac Equation in Kerr-Newman-dS Black Hole Background}

\author{Francesco Belgiorno}
\address{Dipartimento di Fisica,
Universit\`a degli Studi di Milano,  Via Celoria 16, 20133 Milano, Italy}
\email{belgiorno@mi.infn.it}
\author{Sergio L. Cacciatori}
\address{
Dipartimento di Fisica, Universit\`a degli Studi dell'Insubria,
Via Valleggio 11, 22100 Como, Italy}
\email{sergio.cacciatori@uninsubria.it}

%\maketitle

\begin{abstract}
We consider the Dirac equation on the background of a Kerr-Newman-de Sitter black hole.
By performing variable separation, we show that there exists no time-periodic and normalizable
solution of the Dirac equation. This conclusion holds true even in the extremal case.
With respect to previously considered cases, the novelty is represented by the presence, 
together with a black hole event horizon, of a
cosmological (non degenerate) event horizon, 
which is at the root of the possibility to draw 
a conclusion on the aforementioned topic
in a straightforward way even in the extremal case.
\end{abstract}

\maketitle

\section{Introduction}

In this paper we extend the results obtained for the Dirac equation on the background of a
Kerr-Newman-AdS black hole \cite{belgcaccia-ads} to the case of a Kerr-Newman-de Sitter black hole.
The main differences between the AdS and the dS case is the presence of a positive cosmological
constant in the dS case (to be compared with the negative cosmological constant of the AdS case),
the replacement of a boundary-like behavior of infinity in the AdS case with the presence of a
further (non degenerate) event horizon in the dS case: the cosmological horizon appears. 
Problems with the lack of global hyperbolicity disappear and a good behavior of the
wave operators is shown to be allowed in the dS case. From the point of view of quantum field theory on the
given background, with respect to the case of a single event horizon, further difficulties
appear, due to the presence in the non extremal case of two different background temperatures
which make trickier a rigorous analysis. We do not deal with this problem herein, and we
limit ourselves to study the problem of the absence of time-periodic normalizable solutions
of the Dirac equation. The latter topic has given rise to a number of studies in the recent
literature \cite{Finster:2000jz,finster-rn,finster-axi,yamada,schmid,baticschmid,hafner,dafermos},
mostly involved in black holes of the Kerr-Newman family, or still in absence of cosmological
constant. We also considered this problem in the case of Kerr-Newman-AdS black holes \cite{belgcaccia-ads}.
In the aforementioned studies the absence of time-periodic normalizable solutions of the Dirac
equation has been proved mainly in the non-extremal case. The extremal one has been shown to 
require further investigation,
and in the Kerr-Newman case the existence of normalizable time-periodic 
solutions was proved in \cite{schmid,yamada}.\\
It is a peculiar property of the background considered herein to forbid the existence of
time-periodic normalizable solutions for the Dirac equation even in the extremal case, and this can be
proved in a rather straightforward way. Naively, the presence of a cosmological event horizon,
which is surely non degenerate in our setting, does not allow to get normalizability of the solutions
near the cosmological horizon. Moreover, this presence is also at the root of the fact that the reduced
radial Hamiltonian, obtained by variable separation, has an absolutely continuous spectrum
which coincides with $\RR$.

\section{The Kerr-Newman-dS solution.}
\label{knads}
The background geometry underlying our problem arises as follows.
One first solves the Einstein-Maxwell equations with a cosmological constant, and next adds
a Dirac field minimally coupled to the electromagnetic field. The Einstein-Maxwell action is
\eqn
S[g_{\mu\nu}, A_\rho]=-\frac 1{16\pi} \int (R-2\Lambda)\sqrt {-\det g} d^4x -\frac 1{16\pi}\int \frac 14 F_{\mu\nu}F^{\mu\nu}
\sqrt {-\det g} d^4x \ ,
\feqn
where $\Lambda=\frac 3{l^2}$ is the positive cosmological constant, $R$ the scalar curvature and $F_{\mu\nu}$ the field strength associated to
the potential 1--form $A$:
\eqn
&& F=dA \ , \qquad\ F_{\mu\nu}=\partial_\mu A_\nu -\partial_\nu A_\mu \ ;\\
&& R=g^{\mu\nu} R_{\mu\nu}\ , \qquad\
R_{\mu\nu}=\partial_\rho \Gamma^\rho_{\mu\nu} -\partial_\nu \Gamma^\rho_{\mu\rho}+\Gamma^{\sigma}_{\mu\nu} \Gamma^\rho_{\sigma\rho}
-\Gamma^{\sigma}_{\mu\rho} \Gamma^\rho_{\sigma\nu} \ ,\\
&& \Gamma^\mu_{\nu\rho} =\frac 12 g^{\mu\sigma}(\partial_\nu g_{\sigma\rho}+\partial_\rho g_{\sigma\nu}-\partial_\sigma g_{\nu\rho})\ .
\feqn
The equations of motion are
\eqn
&& R_{\mu\nu}-\frac 12 (R-2\Lambda) g_{\mu\nu}= -2 \left(F_\mu^{\ \ \rho} F_{\nu\rho} -\frac 14 g_{\mu\nu} F_{\rho\sigma}F^{\rho\sigma}\right)
\ , \\
&& \partial_\mu (\sqrt {-\det g} F^{\mu\nu} )=0\ .
\feqn
With respect to a set of vierbein one forms
\eqn
e^i= e^i_\mu dx^\mu\ , \qquad\ i=0,1,2,3\ ,
\feqn
we have
\eqn
ds^2 =g= \eta_{ij} e^i\otimes e^j\ , \qquad\ g_{\mu\nu}=\eta_{ij} e^i_\mu e^j_\nu\ ,
\feqn
where $\eta=\diag (-1,1,1,1)$ is the usual flat Minkowski metric, so that, as usual, we define the $so(1,3)$ valued spin connection one
forms $\omega^i_{\ j}$ such that
\eqn
de^i +\omega^i_{\ j}\wedge e^j=0\ .
\feqn
We will consider the following background solution.\\

The metric is (cf. e.g. \cite{carter,dehghani,mann})
\eqn
&& ds^2 =-\frac {\Delta_r}{\rho^2}\left[ dt-\frac {a\sin^2 \theta}\Xi d\phi \right]^2 +\frac {\rho^2}{\Delta_r} dr^2
+\frac {\rho^2}{\nat} d\theta^2 +\nat \frac {\sin^2 \theta}{\rho^2} \left[ a dt -\frac {r^2+a^2}\Xi d\phi \right]^2\ ,
\feqn
where
\eqn
&& \rho^2=r^2+a^2\cos^2 \theta\ , \qquad \Xi =1+\frac {a^2}{l^2}\ , \qquad \nar=(r^2+a^2)\left(1-\frac {r^2}{l^2} \right)-2mr+z^2\ ,\\
&& \nat =1+\frac {a^2}{l^2} \cos^2 \theta \ , \qquad z^2=q_e^2 +q_m^2\ ,
\feqn
and the electromagnetic potential and field strength are
\eqn
&& A=-\frac {q_e r}{\rho \snar} e^0 -\frac {q_m \cos \theta}{\rho \snat \sin \theta} e^1\ ,\\
&& F=-\frac {1}{\rho^4}[q_e (r^2-a^2\cos^2\theta)+2q_m ra \cos \theta]e^0 \wedge e^2\cr
&&\qquad\ +\frac {1}{\rho^4}[q_m (r^2-a^2\cos^2\theta)-2q_e ra \cos \theta]e^3 \wedge e^1\ ,
\feqn
where we introduced the vierbein
\eqn
&& e^0 =\frac {\snar}{\rho} \left( dt-\frac {a\sin^2 \theta}\Xi d\phi \right)\ ,\\
&& e^1 =\frac {\snat \sin \theta}\rho \left( a dt -\frac {r^2+a^2}{\Xi} d\phi  \right)\ , \\
&& e^2 =\frac \rho{\snar} dr \ ,\\
&& e^3 =\frac \rho{\snat} d\theta \ .
\feqn
We are interested in the case where three real positive zeroes of $\nar$ appear: a cosmological
event horizon radius $\rc$, a black hole event horizon $\rp<\rc$, a Cauchy horizon $\rmen \leq \rp$,
with the extremal case which is implemented when $\rmen=\rp$ and the non-extremal case implemented
otherwise.  The following reparameterization of $\nar$ is useful:
\eqn
\Delta_r = \frac{1}{l^2} (\rc-r) (r-\rp) (r-\rmen) (r+\rc+\rp+\rmen),
\label{reparam-delta}
\feqn
where the parameters $m,z^2,a^2,l$ are replaced by $\rc,r_{+},r_{-},l$.
One easily finds:
\begin{eqnarray*} %\eqn
m &=& \frac{1}{2 l^2} (\rc+\rp) (\rc+\rmen) (\rp+\rmen)\\
a^2 &=& l^2 - (\rc^2+\rp^2+\rmen^2+\rc \rp + \rc \rmen + \rp \rmen)\\
z^2 &=& \frac{1}{l^2} r_{c} r_{+} r_{-} (r_{c}+r_{+}+r_-) -a^2.
\end{eqnarray*} %\feqn
We note that the above reparameterization implies $a^2<l^2$. As to the determinant ${\mathcal J}$
of the Jacobian matrix, we find
\eqn
{\mathcal J} &=&-\frac{1}{2 l^4}(\rc-\rp)(\rc-\rmen)(\rp - \rmen)(2\rc+\rp+\rmen)(\rc+2\rp+\rmen)(\rc+\rp+2\rmen)
\feqn
which is negative everywhere in the non-extremal case. In the extremal one,
an analogous reparameterization exists, with the only caveat that the
number of independent parameters is three (e.g. $z^2,a^2,l$).\\
As to the existence of black hole solutions for given values of the geometrical
parameters, a study of the existence of zeroes for $\nar$ is required (see also 
\cite{dehghani,mann}). A first
observation is that, in order that there exist four real zeroes, it is necessary that
$\frac{d^2 \nar}{d r^2}$ admits two real zeroes, and this leads
again to the condition $l^2-a^2>0$. Qualitatively, one can point out that for $m=0$
the function $\nar$ admits only two real zeroes $r_0<\rc$ and two (symmetric) positive maxima and
one positive minimum between them; 
for increasing $m>0$, the minimum eventually intersects the $r$-axis, say at $m=m_{crit}^{-}$, 
providing the existence of two further zeroes $\rmen\leq \rp$ which coincide for 
$m=m_{crit}^{-}$ (extremal black hole); an upper bound $m_{crit}^{+}$
to $m$ has still to be set, because  
the maximum on the right of the minimum eventually reaches the $r$-axis, where $\rp = \rc$, and then for
$m>m_{crit}^{+}$
again two real solutions remain. The aforementioned two critical situations
are obtained by solving the system
\eqn
&&\nar = 0 \\
&&\nar'=0,
\feqn
where $\nar':=\frac{d \nar}{d r}$, 
i.e. the equivalent system 
\eqn
&&\nar - r \nar' = 0 \label{zer}\\
&&\nar' =0  \label{zerp}.
\feqn
Eqn. (\ref{zer}) amounts to
\eqn
3r^4-r^2(l^2-a^2)+l^2(a^2+z^2)=0;
\feqn
its solutions are
\eqn
R_{\pm}=\sqrt{\frac {l^2-a^2}6 \pm \frac 16\sqrt{(l^2-a^2)^2-12l^2(a^2+z^2)}},
\feqn
and their existence, with $R_+>R_{-}$, requires the condition $(l^2-a^2)^2-12l^2(a^2+z^2)>0$,
i.e.
\eqn
\frac {a^2}{l^2} \leq 7-4\sqrt 3,
\label{forte}
\feqn
which is sensibly more restrictive than $\frac {a^2}{l^2}<1$. Then, from (\ref{zerp}),
one finds the corresponding critical values of the mass:
\eqn
m_{crit}^{\pm}=\frac {R_\pm^2}{l^2}[l^2-a^2-2R_{\pm}^2].
\feqn
Then we find the condition (together with (\ref{forte})) to be satisfied:
\eqn
m_{crit}^{-} \leq m < m_{crit}^{+},
\feqn
with
\eqn
m_{crit}^{\pm}&=& \frac {l}{3\sqrt 6}\left(\left( 1-\frac {a^2}{l^2}\right)\pm \sqrt {\left( 1-\frac {a^2}{l^2}\right)^2-\frac {12}{l^2} (a^2+z^2)}
\right)^{\frac 12}\cr
&&\times \left( 2\left( 1-\frac {a^2}{l^2}\right)\mp \sqrt {\left( 1-\frac {a^2}{l^2}\right)^2-\frac {12}{l^2} (a^2+z^2)}\right).
\feqn
The same conditions can be obtained by studying the cubic resolvent associated with the equation $\Delta =0$: 
\eqn
u^3-u\tilde p -\tilde q =0, 
\label{cubic-res}
\feqn
where $\tilde p =4w+\frac {p^2}3,\ \tilde q =\frac {2p^3}{27} +q^2 -\frac {8pw}3$, and where 
$p=-(l^2-a^2),\ q=2ml^2,\ w=-l^2(a^2+z^2)$. The solutions of (\ref{cubic-res}) are all real 
iff
\eqn
\frac {\tilde q^2}{4}-\frac {\tilde p^3}{27}\leq 0,
\feqn
i.e. one has to impose $\tilde p> 0$ (which amounts to (\ref{forte})) and 
\eqn
\left[ 2m^2 l^4 -\frac 43 l^2(l^2-a^2)(a^2+z^2)-\frac 1{27} (l^2-a^2)^3 \right]^2-\frac 1{27} \left[ \frac {(l^2-a^2)^2}3-4l^2(a^2+z^2) \right]^3\leq 0.
\feqn
The above inequality is implemented for $m_{crit}^{-} \leq m \leq m_{crit}^{+}$ 
and for $-m_{crit}^{+} \leq m \leq -m_{crit}^{-}$. The latter solution would correspond to negative values of 
the mass parameter $m$. 
Note also that for $m= m_{crit}^{+}$ one would obtain a black hole with $\rp=\rc$. We do not discuss the latter case herein.

%\newpage

\section{The Dirac equation.}
\label{diraceq}

The Dirac equation for a charged massive particle of mass $\mu$ and electric charge $e$ is
\eqn
(i\gamma^\mu D_\mu -\mu )\psi=0\ ,
\feqn
where $D$ is the Koszul connection on the bundle $S\otimes U(1)$, $S$ being the spin bundle over the 
Kerr-Newman-dS manifold, that is
\eqn
D_\mu =\de_\mu +\frac 14 \omega_\mu^{\ ij} \Gamma_i \Gamma_j +ie A_\mu\ .
\feqn
Here $\omega^{ij}=\omega^i_{\ k} \eta^{kj}$ are the spin connection one forms associated to a vierbein $v^i$, such that $ds^2=\eta_{ij} v^i \otimes v^j$, $\eta$
being the usual Minkowski metric.
$\gamma_\mu$ are the local Dirac matrices, related to the point independent Minkowskian Dirac matrices $\Gamma_i$ by the relations
$\gamma_\mu =v_\mu^i \Gamma_i$.\\
Here we use the representation
\eqn
\Gamma^0 =\left(
\begin{array}{cc}
\mathbb {O} & -\mathbb {I} \\
-\mathbb {I} & \mathbb {O}
\end{array}
\right)\ ,
\qquad
\vec \Gamma =\left(
\begin{array}{cc}
\mathbb {O} & -\vec \sigma \\
\vec \sigma & \mathbb {O}
\end{array}
\right)\ ,
\feqn
where
\eqn
\mathbb {O} =\left(
\begin{array}{cc}
0 & 0 \\
0 & 0
\end{array}
\right)\ ,
\qquad
\mathbb {I} =\left(
\begin{array}{cc}
1 & 0 \\
0 & 1
\end{array}
\right)\ ,
\feqn
and $\vec \sigma$ are the usual Pauli matrices
\eqn
\sigma_1  =\left(
\begin{array}{cc}
0 & 1 \\
1 & 0
\end{array}
\right)\ ,
\qquad
\sigma_2  =\left(
\begin{array}{cc}
0 & -i \\
i & 0
\end{array}
\right)\ ,
\qquad
\sigma_3  =\left(
\begin{array}{cc}
1 & 0 \\
0 & -1
\end{array}
\right)\ .
\feqn
Thus
\eqn
\gamma_\mu \gamma_\nu +\gamma_\nu \gamma_\mu =-2 g_{\mu\nu}\ .
\feqn

Following the general results of \cite{Kamran:1984mb} one can obtain variable separation as in \cite{belgcaccia-ads}.
We limit ourselves to display the final result herein.
The Petrov type D condition ensures the existence of a phase function ${\mathcal B} (r,\theta)$ such that
\eqn
d{\mathcal B} =\frac 1{4Z(r,\theta)} \left( -2a \frac {\cos \theta}\Xi dr -2ar \frac {\sin \theta}\Xi d\theta \right)\ ,
\feqn
which indeed gives
\eqn
{\mathcal B}(r,\theta)=\frac i4 \log \frac {r-ia\cos\theta}{r+ia\cos\theta} \ .
\feqn
Now let us write the Dirac equation as
\eqn
H_D \psi=0\ .
\feqn
Under a transformation $\psi \mapsto S^{-1}\psi$, with
\eqn
S=Z^{-\frac 14}\diag (e^{i{\mathcal B}},e^{i{\mathcal B}},e^{-i{\mathcal B}},e^{-i{\mathcal B}})\ ,
\feqn
it changes as
\eqn
S^{-1} H_D S (S^{-1}\psi)=0\ .
\feqn
If we multiply this equation times
\eqn
U=iZ^{\frac 12}\diag (e^{2i{\mathcal B}},-e^{2i{\mathcal B}},-e^{-2i{\mathcal B}},e^{-2i{\mathcal B}})\ ,
\feqn
and introduce the new wave function
\eqn
\tilde \psi =(\nat \nar)^{\frac 14} S^{-1} \psi\ , \label{reduced}
\feqn
then the Dirac equation takes the form
\eqn
({\mathcal R}(r)+{\mathcal A}(\theta)) \tilde \psi =0\ , \label{dirac}
\feqn
where
\eqn
&& {\mathcal R}=\left(
\begin{array}{cccc}
i\mu r  & 0 & -\snar {\mathcal D}_+ & 0 \\
0 & -i\mu r & 0 & -\snar {\mathcal D}_- \\
-\snar {\mathcal D}_- & 0 & -i\mu r & 0 \\
0 & -\snar {\mathcal D}_+ & 0 & i \mu r
\end{array}
\right) \ , \\
&& {\mathcal A}=\left(
\begin{array}{cccc}
-a\mu \cos \theta  & 0 & 0 & -i\snat {\mathcal L}_- \\
0 & a\mu \cos \theta  & -i\snat {\mathcal L}_+ & 0  \\
0 & -i\snat {\mathcal L}_- & -a\mu \cos \theta & 0  \\
-i\snat {\mathcal L}_+ & 0 & 0 &  a \mu \cos \theta
\end{array}
\right) \ ,
\feqn
and
\eqn
&& {\mathcal D}_{\pm} =\de_r \pm \frac 1{\nar}\left( (r^2+a^2)\de_t -a\Xi \de_\phi +ie q_e r \right)\ ,\\
&& {\mathcal L}_{\pm} =\de_\theta +\frac 12 \cot \theta \pm \frac i{\nat \sin \theta} \left( \Xi \de_\phi -a\sin^2 \theta \de_t
+ie q_m \cos \theta \right) \ .
\feqn
Separation of variables can then be obtained searching for solutions of the form
\eqn\label{separation}
\tilde \psi (t,\phi, r, \theta)=e^{-i\omega t}e^{-ik \phi}
\left(
\begin{array}{c}
R_1 (r) S_2 (\theta)\\
R_2 (r) S_1 (\theta)\\
R_2 (r) S_2 (\theta)\\
R_1 (r) S_1 (\theta)
\end{array}
\right)\ , \qquad k \in \mathbb{Z}+\frac 12 \ .
\feqn

%\newpage
\section{Hamiltonian formulation.}
\label{hamilton}

The Hamiltonian for the Dirac equation can be read from (\ref{dirac}) rewriting it in the form \cite{Finster:2000jz}
\eqn
i\de_t \tilde \psi =H \tilde \psi\ .
\label{hamilton-dirac}
\feqn
Indeed we find
\eqn
H=\left[ \left(1-\frac {\nar}{\nat} \frac {a^2 \sin^2 \theta}{(r^2+a^2)^2} \right)^{-1}
\left( \mathbb {I}_4 -\frac {\snar}{\snat} \frac {a\sin \theta}{r^2 +a^2} B C \right)\right](\tilde{\mathcal R}+\tilde{\mathcal A})\ ,\label{hamiltonian}
\feqn
where $\mathbb {I}_4$ is the $4\times 4$ identity matrix,
\eqn
&&\tilde{\mathcal R}= -\frac {\mu r \snar}{r^2+a^2}
\left(
\begin{array}{cccc}
0 & 0 & 1 & 0 \\
0 & 0 & 0 & 1 \\
1 & 0 & 0 & 0 \\
0 & 1 & 0 & 0
\end{array}
\right)
+
\left(
\begin{array}{cccc}
{\mathcal E}_- & 0 & 0 & 0 \\
0 & -{\mathcal E}_+ & 0 & 0 \\
0 & 0 & -{\mathcal E}_+ & 0 \\
0 & 0 & 0 & {\mathcal E}_-
\end{array}
\right)\ , \\
&& \tilde{\mathcal A}=\frac {a \mu \cos \theta \snar}{r^2+a^2}
\left(
\begin{array}{cccc}
0 & 0 & i & 0 \\
0 & 0 & 0 & i \\
-i & 0 & 0 & 0 \\
0 & -i & 0 & 0
\end{array}
\right)
+
\left(
\begin{array}{cccc}
0 & -{\mathcal M}_- & 0 & 0 \\
{\mathcal M}_+ & 0 & 0 & 0 \\
0 & 0 & 0 & {\mathcal M}_- \\
0 & 0 & -{\mathcal M}_+ & 0
\end{array}
\right)\ ,\\
&& {\mathcal E}_\pm =i \frac {\nar}{a^2+r^2} \left[ \de_r \mp \frac {a\Xi}{\nar} \de_\phi \pm i \frac {e q_e r}{\nar} \right]\ ,\\
&& {\mathcal M}_\pm =\frac {\snar \snat}{r^2+a^2} \left[ \de_\theta +\frac 12 \cot \theta \pm \frac {i\Xi}{\nat \sin \theta} \de_\phi
\mp \frac {e q_m \cot \theta}{\nat} \right]\ ,
\feqn
and
\eqn
B=
\left(
\begin{array}{cccc}
0 & 0 & -i & 0 \\
0 & 0 & 0 & i \\
i & 0 & 0 & 0 \\
0 & -i & 0 & 0
\end{array}
\right)\ , \qquad
C=\left(
\begin{array}{cccc}
0 & 0 & 0 & i \\
0 & 0 & -i & 0 \\
0 & i & 0 & 0 \\
-i & 0 & 0 & 0
\end{array}
\right)
\feqn
satisfy $[B,C]=0$, $B^2=C^2=\mathbb {I}_4$. Cf. also \cite{Finster:2000jz} for the Kerr-Newman case.
We need now to specify the Hilbert space. We do it as follows, in strict analogy with \cite{belgcaccia-ads}. 
If we foliate spacetime in $t=constant$ slices ${\mathcal S}_t$, the metric on any slice (considering the shift vectors) is
\eqn
d\gamma^2 =\gamma_{\alpha \beta} dx^\alpha dx^\beta \ ,
\feqn
where $\alpha=1,2,3$ and
\eqn
\gamma_{\alpha\beta} =g_{\alpha\beta}-\frac {g_{0\alpha} g_{0\beta}}{g_{00}} \ ,
\feqn
and local measure
\eqn
d\mu_3= \sqrt {\det \gamma}\ dr d\theta d\phi =\frac {\sin \theta}{\Xi} \frac {\rho^3}{\sqrt {\nar -a^2 \nat \sin^2 \theta}}
\ dr d\theta d\phi \ . \label{measure3}
\feqn
In particular the four dimensional measure factors as
\eqn
d\mu_4= \sqrt {-g_{00}} d\mu_3 dt \ . \label{factor}
\feqn
The action for a massless uncharged Dirac particle is then
\eqn
S=\int_{\mathbb{R}} dt \int_{{\mathcal S}_t} \sqrt {-g_{00}}\ ^t \psi^* \Gamma^0 \gamma^\mu D_\mu \psi d\mu_3 \ ,
\feqn
where the star indicates complex conjugation. Here with ${\mathcal S}_t$ we mean the range of coordinates parameterizing the region external to the
event horizon: $r>r_+$, that is ${\mathcal S}_t:= {\mathcal S} =(r_+, \rc)\times (0,\pi)\times (0,2\pi)$.
Then, the scalar product between wave functions should be
\eqn
\langle \psi | \chi \rangle = \int_{{\mathcal S}} \sqrt {-g_{00}}\ ^t \psi^* \Gamma^0 \gamma^t \chi d\mu_3 \ .
\feqn
We can now use (\ref{measure3}), (\ref{reduced}) and the relation
\eqn
\gamma^2=e^t_0 \Gamma^0 +e^t_1 \Gamma^1\ ,
\feqn
to express the product in the space of reduced wave functions (i.e. (\ref{reduced})):
\eqn
\langle \tilde \psi |\tilde \chi \rangle =\int_{r_+}^{\rc} dr \int_0^\pi d\theta \int_0^{2\pi} d\phi \frac {r^2+a^2}\nar
\frac {\sin \theta}{\snat}\ ^t\tilde \psi^* \left(\mathbb {I}_4 +\frac {\snar}{\snat} \frac {a\sin \theta}{r^2 +a^2} B C \right)\tilde \chi \ ,
\label{product}
\feqn
where a factor $\Xi^{-\frac 12}$ has been dropped. The matrix in the parenthesis in the previous equation 
is the inverse of the one 
in the square brackets in (\ref{hamiltonian}), and it represents an improvement to the dS case of 
the matrix which has been introduced in \cite{Finster:2000jz} for the Kerr-Newman case.\\
The above scalar product is positive definite, as we show in the following. 
Being $\pm 1$ the eigenvalues of $BC$, we need to prove that
\eqn
\eta:=\sup_{r\in (\rp,\rc), \theta\in (0,\pi)} \alpha(r,\theta)<1\ ,
\feqn
where
\eqn
\alpha(r,\theta)=\frac {\snar}{\snat} \frac {a\sin\theta}{r^2+a^2}\ .
\feqn
We can write $\alpha(r,\theta)=\beta(r)\gamma(\theta)$, with
\eqn
\gamma(\theta)=\frac {\sin\theta}{\snat}  \ .
\feqn
Then
\eqn
\gamma'(\theta)=\frac {\cos\theta}{\nat^{\frac 32}}\left( 1+\frac {a^2}{l^2} \right)
\feqn
so that $\gamma$ reaches its maximum at $\theta=\pi/2$ and
\eqn
\gamma(\pi/2)=1\ ,
\feqn
which implies
\eqn
\alpha (r,\theta)\leq \beta(r).
\feqn
Next, from
\eqn
0=\nar(r_+)
\feqn
we have
\eqn
z^2-2mr_+=-(r_+^2+a^2)(l^2-r^2)/l^2 <0\ ,
\feqn
and then, for $\rp \leq r \leq \rc$ we have $z^2-2mr<0$ (note that $l^2> \rc^2$ for our case).
Thus
\eqn
\beta^2 (r)=\frac {a^2\nar}{(r^2+a^2)^2}=\frac {a^2}{l^2}\frac {l^2-r^2}{r^2+a^2}+a^2\frac {z^2-2mr}{(r^2+a^2)^2}\leq
\frac {a^2}{l^2}\frac {l^2-r^2}{r^2+a^2} =: h(r)\ .
\feqn
Now, the last function is a decreasing function of $r$, so that for $r\geq r_+>0$ we have $h(\rc) \leq h(r)\leq h(r_+)< h(0)$, so that
\eqn
\beta^2 (r)\leq h(r_+)=\frac {a^2}{l^2}\frac {l^2-\rp^2}{r_+^2+a^2}<h(0)=1\ ,
\feqn
and then
\eqn
\eta \leq \sqrt {h(r_+)} <1\ .
\feqn

\section{Essential selfadjointness of \boldmath{$\hat H$}.}
\label{essauto}

We follow strictly our analysis in \cite{belgcaccia-ads}, limiting ourselves to some essential definitions 
and results. Let us introduce the space of functions ${\mathcal L}^2:=(L^2((r_+,\rc) \times S^2; d\mu))^4$ with
measure
\eqn
d\mu = \frac {r^2+a^2}\nar \frac {\sin \theta}\snat dr d\theta d\phi.
\feqn
and define ${\cHH}_{<>}$ as the Hilbert space ${\mathcal L}^2$ with the scalar product
(\ref{product}). We will also consider a second Hilbert space
$\cHH_{()}$, which is obtained from  ${\mathcal L}^2$ with the scalar product
\eqn
( \psi | \chi ) =\int_{r_+}^{\rc} dr \int_0^\pi d\theta \int_0^{2\pi} d\phi \frac {r^2+a^2}\nar
\frac {\sin \theta}{\snat}\ ^t \psi^* \chi \ =\int d\mu ^t\psi^* \chi \ .
\label{ri-product}
\feqn
It is straightforward to show that $||\cdot||_{<>}$ and $||\cdot||_{()}$ are equivalent norms.
It is also useful to introduce $\hat \Omega^2:{\mathcal L}^2\to {\mathcal L}^2$ as the multiplication
operator by $\Omega^2 (r,\theta)$:
\eqn
\Omega^2 (r,\theta):=\mathbb {I}_4 +\alpha (r,\theta)\;  B C\ . \label{omega2}
\feqn
Then we have
\eqn
\langle \psi | \chi \rangle =\int d\mu ^t\psi^* \Omega^2 \chi \ = ( \psi | \hat \Omega^2 \chi ) \ .
\label{red_prod}
\feqn
We introduce also $\hat \Omega^{-2}:{\mathcal L}^2\to {\mathcal L}^2$ as the multiplication operator by $\Omega^{-2}$:
\eqn
\Omega^{-2} (r,\theta):=\frac{1}{1-\alpha^2 (r,\theta)}\left(\mathbb {I}_4 -\alpha (r,\theta)\;
B C \right)\ ,  \label{omega-2}
\feqn
and analogously $\hat \Omega,\hat \Omega^{-1}$ are defined as operators from
${\mathcal L}^2$ to ${\mathcal L}^2$ which multiply by $\Omega (r,\theta)$,
$\Omega^{-1} (r,\theta)$ respectively, where $\Omega$ and $\Omega^{-1}$ are defined as
the principal square root of $\Omega^2$ and $\Omega^{-2}$ respectively. They are 
injective and surjective. As operators from $\cHH_{()}$ to $\cHH_{()}$, 
$\hat \Omega^2, \hat \Omega^{-2}, \hat \Omega,\hat \Omega^{-1}$ are 
bounded, positive and selfadjoint. 

Let us set $H_0 :=\tilde{\mathcal R}+\tilde{\mathcal A}$, which is formally selfadjoint on $\cHH_{()}$,
and define the operator $\hat  H_0$ on ${\mathcal L}^2$
with
\eqn
&&D(\hat H_0)= C_0^{\infty} ((r_+,\rc) \times S^2)^4 =:\cDD \cr
&&\hat H_0 \chi = H_0 \chi, \quad \chi \in \cDD.
\feqn
Notice that $\cDD$ is dense in ${\cHH}_{()}$.
Let us point out that for the formal differential expression $H$ in (\ref{hamiltonian}),
which is formally selfadjoint on $\cHH_{<>}$,
one can write $H=\Omega^{-2} H_0$,
Then we define on ${\mathcal L}^2$
the differential operator $\hat H= \hat \Omega^{-2} \hat H_0$, with
\eqn
&&D(\hat H) = \cDD\cr
&&\hat H \chi = H \chi, \quad \chi \in \cDD.
\label{h-h0}
\feqn
The same considerations as in \cite{belgcaccia-ads} lead to the following conclusions:
$\hat H$ is essentially self-adjoint if and only if $\hat H_0$ is essentially self-adjoint
on the same domain (in different Hilbert spaces). Cf. Theorem 1 in \cite{belgcaccia-ads}. 
See also \cite{winklyamada} for the Kerr-Newman case. 
Moreover, one can show by means of variable separation (cf. \cite{belgcaccia-ads}) 
that $\hat H_0$ is essentially self-adjoint. As a consequence, there exists a unique
self-adjoint extension $\hat T_{H_0}$ with domain ${\mathfrak D}\subset \cHH_{()}$ and,
correspondingly, a unique self-adjoint extension  $\hat T_{H}:=\hat \Omega^{-2} \hat T_{H_0}$ of $\hat H$ on
${\mathfrak D}\subset \cHH_{<>}$ (note that it is the same domain on two different Hilbert spaces).
We do not report the details of the variable separation process, because they are the same as in
\cite{belgcaccia-ads}. We limit ourselves to sketch the main points. By means of the unitary transformation 
\eqn
V=
\frac{1}{\sqrt{2}}
\left(
\begin{array}{cccc}
 0  & -i &  0  &  i \\
 i  &  0 & -i  &  0 \\
 0  & -1 &  0  & -1 \\
-1  &  0 & -1  &  0
\end{array}
\right)
\label{def-V}
\feqn
we get 
\eqn
V H_0  V^{\ast}=
\left(
\begin{array}{cc}
\frac{1}{r^2+a^2} (i a \Xi \partial_{\phi} +e q_e r +\mu r \sqrt{\Delta_r}) {\mathbb {I}}
& \frac{\Delta_r}{r^2+a^2} \partial_r {\mathbb {I}} + \frac{\sqrt{\Delta_r}}{r^2+a^2} {\mathbb {U}}\\
-\frac{\Delta_r}{r^2+a^2} \partial_r {\mathbb {I}} + \frac{\sqrt{\Delta_r}}{r^2+a^2} {\mathbb {U}}&
\frac{1}{r^2+a^2} (i a \Xi \partial_{\phi} +e q_e r -\mu r \sqrt{\Delta_r}) {\mathbb {I}}
\end{array}
\right), 
\label{red-h0}
\feqn
where ${\mathbb {U}}$ is the $2\times 2$ matrix formal differential expression 
\eqn
{\mathbb {U}}=
\left(
\begin{array}{cc}
-\mu a \cos (\theta) & i \sqrt{\Delta_{\theta}} (\partial_{\theta}+\frac{1}{2} \cot ({\theta})
+ g)\\
%i \frac{1}{\Delta_{\theta} \sin (\theta)} \Xi \partial_{\phi} -\frac{1}{\Delta_{\theta}} q_m e \cot ({\theta}) )\\
 i \sqrt{\Delta_{\theta}} (\partial_{\theta}+\frac{1}{2} \cot ({\theta})
- g)) &
%-i \frac{1}{\Delta_{\theta} \sin (\theta)} \Xi \partial_{\phi} +\frac{1}{\Delta_{\theta}} q_m e \cot ({\theta}) )&
\mu a \cos (\theta)
\end{array}
\right),
\feqn
with $g:=i \frac{1}{\Delta_{\theta} \sin (\theta)} \Xi \partial_{\phi}
-\frac{1}{\Delta_{\theta}} q_m e \cot ({\theta})$.\\ 
Then the following variable separation ansatz 
\eqn\label{separa-h}
V \chi (r,\theta,\phi)=\frac{e^{-i k \phi}}{\sqrt{2\pi}}  V 
\left(
\begin{array}{c}
R_1 (r) S_2 (\theta)\\
R_2 (r) S_1 (\theta)\\
R_2 (r) S_2 (\theta)\\
R_1 (r) S_1 (\theta)
\end{array}
\right)\ , 
\feqn
with $k\in {\mathbb Z}+\frac{1}{2}$, leads to the following reduction of the angular part: 
by defining $b_k (\theta):= \frac{1}{\Delta_{\theta} \sin (\theta)} \Xi k
-\frac{1}{\Delta_{\theta}} q_m e \cot ({\theta})$, one finds that the operator  $\hat {\mathbb {U}}_k$ 
defined on $D(\hat {\mathbb {U}}_k)=C_0^{\infty} (0,\pi)^2$ and whose formal 
differential expression is
\eqn
{\mathbb {U}}_k=
\left(
\begin{array}{cc}
-\mu a \cos (\theta) & i \sqrt{\Delta_{\theta}} (\partial_{\theta}+\frac{1}{2} \cot ({\theta})+ b_k (\theta))\\
%\frac{1}{\Delta_{\theta} \sin (\theta)} \Xi k -\frac{1}{\Delta_{\theta}} q_m e \cot ({\theta}) )\\
 i \sqrt{\Delta_{\theta}} (\partial_{\theta}+\frac{1}{2} \cot ({\theta})
-  b_k (\theta)) &
%- \frac{1}{\Delta_{\theta} \sin (\theta)} \Xi k +\frac{1}{\Delta_{\theta}} q_m e \cot ({\theta}) )&
\mu a \cos (\theta)
\end{array}
\right),
\feqn
is essentially selfadjoint 
for any $k\in {\mathbb Z}+\frac{1}{2}$ for $\frac{q_m e}{\Xi} \in \mathbb{Z}$.
If one considers the selfadjoint extension $\bar{\hat {\mathbb {U}}}_k$ of $\hat {\mathbb {U}}_k$, one can show that
$\bar{\hat {\mathbb {U}}}_k$ has purely discrete spectrum which is simple (see \cite{belgcaccia-ads}).

Let us introduce
the (normalized) eigenfunctions $S_{k;j} (\theta):=\left(
\begin{array}{c}
S_{1\; k;j} (\theta)\\
S_{2\; k;j} (\theta)
\end{array}
\right)
$
of the operator $\bar{\hat {\mathbb {U}}}_k$:
\eqn
\bar{\hat {\mathbb {U}}}_k
\left(
\begin{array}{c}
S_{1\; k;j} (\theta)\\
S_{2\; k;j} (\theta)
\end{array}
\right)
=
\lambda_{k;j}
\left(
\begin{array}{c}
S_{1 \; k;j} (\theta)\\
S_{2 \; k;j} (\theta)
\end{array}
\right),
\label{red-angular}
\feqn
then ${\mathcal H}_{k,j}:=L^2((r_{+},\rc), \frac{r^2+a^2}{\Delta_r} dr)^2\otimes M_{k,j}$, where $M_{k,j}:=\{F_{k;j}(\theta,\phi)\}$,
with $F_{k;j}(\theta,\phi):=S_{k;j}(\theta) \frac{e^{-i k \phi}}{\sqrt{2\pi}}$,
is such that the eigenvalue equation for $V \hat H_0  V^{\ast}$ becomes equivalent to the following $2\times 2$ Dirac system for the radial part (cf. \cite{belgcaccia-ads}):
\eqn
\left(
\begin{array}{cc}
\frac{1}{r^2+a^2} (a \Xi k +e q_e r +\mu r \sqrt{\Delta_r})
& \frac{\Delta_r}{r^2+a^2} \partial_r  + \frac{\sqrt{\Delta_r}}{r^2+a^2} \lambda_{k;j}\\
-\frac{\Delta_r}{r^2+a^2} \partial_r  + \frac{\sqrt{\Delta_r}}{r^2+a^2} \lambda_{k;j} &
\frac{1}{r^2+a^2} (a \Xi k +e q_e r -\mu r \sqrt{\Delta_r})
\end{array}
\right)
\left(
\begin{array}{c}
X_1 (r) \\
X_2 (r)
\end{array}
\right)
=
\omega
\left(
\begin{array}{c}
X_1 (r) \\
X_2 (r)
\end{array}
\right),
\label{reduced-radial}
\feqn
and we introduce a radial Hamiltonian $\hat h_{k,j}$, which is defined on
${\mathcal D}_{k,j}:=C_0^{\infty} (r_+,\rc)^2$ and has the following formal expression:
\eqn
h_{k,j}:=
\left(
\begin{array}{cc}
\frac{1}{r^2+a^2} (a \Xi k +e q_e r +\mu r \sqrt{\Delta_r})
& \frac{\Delta_r}{r^2+a^2} \partial_r  + \frac{\sqrt{\Delta_r}}{r^2+a^2} \lambda_{k;j}\\
-\frac{\Delta_r}{r^2+a^2} \partial_r  + \frac{\sqrt{\Delta_r}}{r^2+a^2} \lambda_{k;j} &
\frac{1}{r^2+a^2} (a \Xi k +e q_e r -\mu r \sqrt{\Delta_r})
\end{array}
\right)
\label{reduction}
\feqn
In the following,
we study essential selfadjointness conditions for the reduced Hamiltonian $\hat h_{k,j}$.\\

%\newpage

\noindent
{\sl Essential selfadjointness of $\hat h_{k,j}$.} 
The differential expression $h_{k,j}$ is formally selfadjoint in the Hilbert space
$L^2 ((r_{+},\rc), \frac{r^2+a^2}{\Delta_r} dr)^2$.
In order to study the essential selfadjointness of the reduced Hamiltonian in $C^{\infty}_0 (r_{+},\rc)^2$ one
has to check if the limit point case occurs both at the event horizon $r=r_{+}$ and at $r=\rc$.
We show that the following result holds:
\begin{theorem}
$\hat h_{k,j}$ is essentially selfadjoint on $C^{\infty}_0 (r_{+},\rc)^2$.
\end{theorem}
\begin{proof}
We choose the tortoise coordinate $y$ defined by
\eqn
dy = \frac{r^2+a^2}{\Delta_r} dr
\label{ytortoise}
\feqn
and obtain $y\in \RR$ with $y\to \infty$ as $r\to \rc$ and $y\to -\infty$ as $r\to \rp$.  Then we get
\eqn
h_{k,j}=
\left(
\begin{array}{cc}
0 & \partial_y\\
-\partial_y & 0
\end{array}
\right)
+ V(r(y)),
\label{hamilton-tortoise}
\feqn
and the corollary to thm. 6.8 p.99 in \cite{weidmann} ensures that the limit point case holds for $h_{k,j}$
at $y=\infty$. The same corollary can be used also for concluding that the limit point case occurs
also for $y=-\infty$ and this allows us to claim that the above theorem holds true.
\end{proof}
It is also useful to point out that it holds
\eqn
\lim_{y\to -\infty} V(r(y))=
\left(
\begin{array}{cc}
\varphi_+ & 0\\
0 & \varphi_+
\end{array}
\right),
\feqn
where
\eqn
\varphi_+ := \frac{1}{r_{+}^2+a^2} (a k  \Xi+ e q_e r_{+}),
\feqn
and that
\eqn
\lim_{y\to \infty} V(r(y))=
\left(
\begin{array}{cc}
\varphi_c & 0\\
0 & \varphi_c
\end{array}
\right),
\feqn
where
\eqn
\varphi_c := \frac{1}{r_{c}^2+a^2} (a k  \Xi+ e q_e r_{c}).
\feqn

\section{The non-existence of time-periodic normalizable solutions.}
\label{eigen-equation}

As it is well-known from the study of the Kerr-Newman case and of the Kerr-Newman-AdS case
\cite{winklmeierthesis,yamada,schmid,baticschmid,belgcaccia-ads}, eigenvalues for the Hamiltonian $H$
correspond to the solutions of the following system
of coupled eigenvalue equations have to be satisfied simultaneously in $L^2 ((0,\pi), \frac{\sin (\theta)}{\sqrt{\Delta_{\theta}}} d\theta)^2$
and in $L^2 ((r_{+},\rc), \frac{r^2+a^2}{\Delta_r} dr)^2$ respectively:
\eqn
\bar{\hat {\mathbb {U}}}_{k\; \omega} S = \lambda S,
\label{angularequation}
\feqn
and
\eqn
\bar{\hat h}_{k,j} X = \omega X.
\label{radialequation}
\feqn
Note that the Dirac equation (\ref{dirac}) in the Chandrasekhar-like variable separation ansatz (\ref{separation})
reduces to the couple of equations (\ref{angularequation}) and (\ref{radialequation}).\\

The spectrum of the angular momentum operator
$\bar{\hat {\mathbb {U}}}_{k\; \omega}$ is discrete for any $\omega\in \mathbb{R}$, as it can be
shown in a step-by-step replication of the calculations appearing in \cite{belgcaccia-ads}. We show that
both in the non-extremal case and in the extremal one
the radial Hamiltonian $\bar{\hat h}_{k,j}$ for any $\lambda_{k;j}$ has a
spectrum is absolutely continuous and coincides with $\mathbb{R}$, and then we infer that
no eigenvalue of $\bar{\hat{H}}$ exists. As a consequence (cf. Remark 1 in \cite{belgcaccia-ads}),
we can exclude the existence of normalizable time-periodic solutions of the Dirac equation.\\

%\newpage
\noindent
{\sl Spectrum of the operator $\bar{\hat h}_{k,j}$.} 
In order to study the spectral properties of $\bar{\hat h}_{k,j}$, as in \cite{belgcaccia-ads} we introduce
two auxiliary selfadjoint operators $\hat h_{hor}$ and $\hat h_{\rc}$:
\eqn
&&D(\hat h_{hor})=\{ X\in L^2_{(r_{+},r_0)},\; X
\hbox{ is locally absolutely continuous}; B(X)=0;\;
%\sin (\beta)X_1 (r_0) + \cos (\beta) X_2 (r_0) =0;\;
\hat h_{hor} X \in L^2_{(r_{+},r_0)}\},\cr
&& \hat h_{hor} X = h_{k,j} X;\cr
&&D(\hat h_{\rc})\hphantom{o}=\{ X\in L^2_{(r_0,\rc)},\; X
\hbox{ is locally absolutely continuous};  B(X)=0;\;
%\sin (\beta) X_1 (r_0)\; +\;\cos (\beta) X_2 (r_0) =0;
\hat h_{\rc} X \in L^2_{(r_0,\rc)}\}\cr
&& \hat h_{\rc} X = h_{k,j} X.
\feqn
$r_0$ is an arbitrary point with $r_{+}<r_0<\rc$, at which the boundary condition $B(X):=
X_1 (r_0) =0$ is imposed, with $X(r):=\left(\begin{array}{c} X_1 (r)\\
X_2 (r)\end{array}\right)$. We also have defined $L^2_{(r_{+},r_0)}:=
L^2 ((r_{+},r_0), \frac{r^2+a^2}{\Delta_r} dr)^2$ and
$L^2_{(r_0,\rc)}:=L^2 ((r_0,\rc), \frac{r^2+a^2}{\Delta_r} dr)^2$.
Note that we omit the indices $k,j$ for these operators.\\

As to the spectral properties of $\hat h_{\rc}$, a suitable change of coordinates consists in
introducing a tortoise-like coordinate defined by eqn. (\ref{ytortoise}).
It is then easy to show that the following result holds.
%\newpage
\begin{lemma}
$\sigma_{ac} (\hat h_{\rc}) = {\mathbb R}$. 
\end{lemma}
\begin{proof}
The proof is completely analogous to the one of Lemma 3 in \cite{belgcaccia-ads}. We still 
provide the details. 
Theorem 16.7 of \cite{weidmann} allows to find that
the spectrum of $\hat h_{\rc}$ is absolutely continuous in ${\mathbb R}-\{\varphi_c\}$.
This can be proved as follows. Let us write the potential $V(r(y))$ in (\ref{hamilton-tortoise})
\eqn
V(r(y))=\left(
\begin{array}{cc}
\varphi_c  & 0\cr
0 & \varphi_c
\end{array}
\right)+P_2 (r(y)),
\label{p1p2}
\feqn
which implicitly defines $P_2 (r(y))$. The first term on the left of (\ref{p1p2}) is of course of
bounded variation; on the other hand, $|P_2 (r(y))|\in L^1 (d,\infty)$, with $d\in (y(r_0),\infty)$. As
a consequence, the hypotheses of theorem 16.7 in \cite{weidmann} are trivially satisfied, and one
finds that the spectrum of $\hat h_{hor}$ is absolutely continuous in  ${\mathbb R}-\{\varphi_c\}$.\\
We show also that $\varphi_c$ is not an eigenvalue of $\hat h_{\rc}$.
As in the Kerr-Newman case (cf. \cite{yamada}), one needs simply to
study the asymptotic behavior of the solutions of the linear system
\eqn
X' =
\left(
\begin{array}{cc}
-\lambda_{k;j} \frac{\sqrt{\Delta_r}}{r^2+a^2}
& \varphi_c - \frac{1}{r^2+a^2} (a \Xi k +e q_e r -\mu r \sqrt{\Delta_r})\\
\frac{1}{r^2+a^2} (a \Xi k +e q_e r +\mu r \sqrt{\Delta_r})-\varphi_c  &
\lambda_{k;j} \frac{\sqrt{\Delta_r}}{r^2+a^2}
\end{array}
\right) X =:\bar{R}(r(y)) X,
\label{eqradialphi}
\feqn
where $r=r(y)$ and where the prime indicates the derivative with respect to $y$.
One easily realizes that 
\eqn
\int_d^{\infty} dy |\bar{R}(r(y))|<\infty,
\feqn
and then (cf. Levinson theorem  e.g. in \cite{eastham}: Theorem 1.3.1 p.8)
one can find two linearly independent 
asymptotic solutions as $y\to \infty$ whose leading order is given by $X_I =\left(
\begin{array}{c} 1\\ 0 \end{array} \right)$ and $X_{II} =\left(
\begin{array}{c} 0\\ 1 \end{array} \right)$. As a consequence no normalizable solution of the
equation (\ref{eqradialphi}) can exists, and then $\varphi_c$ cannot be an eigenvalue.
\end{proof}
The following result holds:
\begin{theorem}
$\sigma_{ac} (\bar{\hat h}_{k,j}) = {\mathbb R}$. 
\end{theorem}
\begin{proof}
Thanks to standard decomposition methods for the absolutely continuous spectrum (see the appendix) 
the proof is trivial, because the absolutely continuous part of $\bar{\hat h}_{k,j}$ is unitarily equivalent to the
absolutely continuous part of $\hat h_{hor}\oplus \hat h_{\rc}$. As a consequence,
$\sigma_{ac} (\bar{\hat h}_{k,j})= \sigma_{ac} (\hat h_{hor})\cup \sigma_{ac} (\hat h_{\rc})$.
The latter set is $\RR$ (see Lemma 1).
\end{proof}

The presence of a cosmological horizon which is non-degenerate (i.e. it corresponds to a simple zero of $\Delta_r$)
is as seen the main ingredient for the above conclusion. A rough explanation for this result is that such a presence
forbids the possibility to get normalizability of the time-periodic solutions. The rationale beyond it is
the above result concerning the absolutely continuous spectrum.

\section{Conclusions}

By extending the results obtained in \cite{belgcaccia-ads}, we have shown that the Dirac Hamiltonian
for a charged particle in the background of a Kerr-Newman-de Sitter black hole is
essentially self-adjoint on $C_0^{\infty} ((\rp,\rc)\times S^2)^4$. Moreover,
the point spectrum of the Hamiltonian is empty, which is equivalent to the condition for absence of
normalizable time-periodic solutions of the Dirac equation. The latter result has been shown to hold true even
in the extremal case, which is usually much harder to be checked, in a rather straightforward way,
and the role of the (non-degenerate) cosmological event horizon in ensuring such a validity has been
pointed out.

%\section*{{\bf Acknowledgments}}
\appendix
\section{Decomposition method for the absolutely continuous spectrum}

For the sake of completeness, we give some more detail about the decomposition method (or splitting method) 
\cite{weidmann,glazman,naimark,ag} as 
applied in the analysis of the absolutely continuous spectrum. It is surely known to experts, but perhaps 
not so explicitly written in the literature. The following proof is essentially 
an extended version, trivially adapted to the Dirac case, of the proof appearing at p.239 of \cite{weidmann} for 
the Sturm-Liouville case, and it also appeals to the proof of Korollar 6.2 in \cite{weidosz}.\\
Let us consider a Dirac system  
with formal differential expression $\tau$, formally selfadjoint in a suitable Hilbert 
space which we indicate with $L_2 (a,b)$ for short, with $(a,b)\subset {\mathbb R}$. 
See \cite{weidmann,weidosz} for more details. 
Let us introduce the maximal operator $\hat K$ 
associated with the formal expression $\tau$, with domain\\ 
$D(\hat K) = \{ X\in L_2 (a,b) ,\; X
\hbox{ is locally absolutely continuous}; \hat K X \in  L_2 (a,b) \}$\\ 
and the minimal operator 
$\hat K_0$ defined as the closure of the operator $\hat K'_0$ defined on\\ 
$D(\hat K'_0) = \{ X\in D(\hat K); X
\hbox{ has compact support in}\ (a,b)\}$ and with formal expression $\tau$. For an explicit 
characterization of $\hat K_0$ see also \cite{weidmann}. 
Let $\hat T$ be a self-adjoint extension 
of $\hat K_0$.\\ 
Let us also define (cf. Korollar 6.2 in \cite{weidosz}) $\hat K_{00}$ as the operator with the 
same formal expression $\tau$ and domain $D(\hat K_{00})=\{ X\in D(\hat K_0); X(c)=0\}$, with $c\in (a,b)$. 
If $\hat K_{a,0},\hat K_{b,0}$ are the minimal operators associated with $\tau$ in 
$L_2 (a,c)$ and $L_2 (c,b)$ respectively, one has 
$\hat K_{00}=\hat K_{a,0}\oplus \hat K_{b,0}$. Let $\hat T_a$ and $\hat T_b$ be self-adjoint extensions of  
$\hat K_{a,0},\hat K_{b,0}$; then both $\hat T$ and 
$\hat T_a \oplus \hat T_b$ are finite-dimensional extensions of $\hat K_{00}$. (Incidentally, 
this is enough for concluding that the essential spectrum of $\hat T$ coincides with the 
essential spectrum of $\hat T_a \oplus \hat T_b$, which is part of the content of Korollar 6.2 
in \cite{weidosz}, and proves the splitting method for the essential spectrum). 
As a consequence, 
the difference of their resolvents $\hat D: = (\hat T-\zeta I)^{-1} -(\hat T_a \oplus \hat T_b -\zeta I)^{-1}$, 
(with $\zeta \in \rho ( \hat T)\cap \rho ( \hat T_a \oplus \hat T_b)$),    
is an operator of finite rank (see \cite{birman}, Lemma 2 p. 214). Then, 
according to the Kuroda-Birman theorem (see e.g. Theorem XI.9 p.27 in \cite{RSIII}; see also \cite{blank} ), the wave operators 
$\Omega^{\pm} (\hat T,\hat T_a \oplus \hat T_b)$ exist and are complete. As a consequence, the absolutely continuous 
part of $\hat T_a \oplus \hat T_b$ is unitarily equivalent to the absolutely continuous part of $\hat T$, 
and this in turn implies that $\sigma_{ac}(\hat T)=\sigma_{ac}(\hat T_a\oplus \hat T_b)=
\sigma_{ac}(\hat T_a)\cup \sigma_{ac}(\hat T_b)$.

%\newpage

\end{document}